# Requirements are All You Need:
# The Final Frontier for End-User Software Engineering

Diana Robinson
University of Cambridge
United Kingdom

Christian Cabrera
University of Cambridge
United Kingdom

Andrew D. Gordon
Cogna and University of Edinburgh
United Kingdom

Neil D. Lawrence
University of Cambridge
United Kingdom

Lars Mennen
Cogna
United Kingdom

## ABSTRACT

What if end users could own the software development lifecycle from conception to deployment using only requirements expressed in language, images, video or audio? We explore this idea, building on the capabilities that generative Artificial Intelligence brings to software generation and maintenance techniques. How could designing software in this way better serve end users? What are the implications of this process for the future of end-user software engineering and the software development lifecycle? We discuss the research needed to bridge the gap between where we are today and these imagined systems of the future.

## 1 INTRODUCTION

What if end users could own the whole software development life cycle from conception through deployment using only *natural requirements*, that is, a mix of natural language, pictures, audio, or even a video demonstration?

End-user software engineering (EUSE) [8, 26] aims to help end-user programmers [4, 37] develop programs more systematically and with high quality. A core challenge is that end users who write and develop programs tend to have no training in or particular interest in software engineering.

Large Language Models (LLMs) have shown impressive capabilities for planning and reasoning, which can support several areas of software engineering [1, 10, 19, 39]. With advances in LLMs, such as OpenAI's GPT-4 [38], and program synthesis, synthesising a whole app directly from user requirements is becoming possible. GitHub's Copilot, launched in mid-2021, empowers professional developers by using an LLM to generate snippets of code from textual prompts, to explain code, to diagnose errors, and to repair them [59]. Consider *whole apps*, that is, pieces of deployable code such as a native desktop or phone app, or a modern website. Since the launch of GPT-4 in 2023, the generation of whole apps from simple natural language requirements has become an active research area. Although today's LLMs struggle to generate whole apps reliably in one go, several research projects aim to synthesise whole apps by chaining calls to LLMs repeatedly and employing LLMs to play different roles such as requirements engineer, software architect, software engineer, or software tester (see for example [22, 53, 53, 58]). Commercial products include Cogna,[1] while there are several open-source projects such as AutoGPT,[2] GPT-Engineer,[3] GPT Pilot,[4] and Devika.[5]

Our vision is that by 2030 end users will build and deploy whole apps just from natural requirements. We call this *requirements-driven end-user software engineering*. New human-AI interfaces will help elicit requirements from the end user, and help them attend to quality and deployment issues for LLM-generated apps. This will transform end users' ability to build software, despite their indifference to engineering. Today, an end user would need a professional engineering team to build a custom app just from requirements. By 2030, relying instead on AI, they can build dramatically more custom apps than now. We expect the area to begin with relatively simple apps, and gradually to scale up, using expert human oversight to complement AI, especially if defects would be consequential.

In what ways will this opportunity matter to end-users? Consider one area, the need to automate data management tasks as the availability of data grows [29]. End-user programming tools marketed as "low code" (since 2014 [6]) or "robotic process automation" (since 2015 [29]) aim to address this need. End users can use these tools to manage data in several domains like e-commerce, business process management, social media, customer management, and content management [34]. However, many of them are prevented by usability aspects [34]. Requirements-driven EUSE will empower many more end users to automate tasks. New AI capabilities will allow more kinds of tasks to be automated.

We need to address the following three challenges:

(1) **Assist end users in expressing their requirements.**

---

[1] *Cogna: Hyper-customised software defined by you, delivered by AI*, February 2024.
[2] *AutoGPT: build & use AI agents*, April 2023.
[3] *GPT-Engineer*, June 2023.
[4] *GPT Pilot*, December 2023.
[5] *Devika is an Agentic AI Software Engineer*, March 2024.

With the flexibility of natural language comes a challenge in guiding users to express their requirements for software precisely and logically enough that they can be built. Users also need to develop an understanding of the potential of software and areas of their workflow that could be automated versus those that could not. There is a huge opportunity here in enabling users to directly influence the creation of software rather than via intermediaries like user researchers and software engineers. This might enable a more nuanced and complete realisation of their goals in software.

There is also scope to support requirements from different end user roles, even within the same software application. We see an opportunity in enabling EUSE in teams where multiple stakeholder roles can bring requirements from their own domain-specific perspectives. Capturing requirements in natural language drops the barrier to participation by providing a shared language to discuss how requirements fit together and possible trade-offs between them at the requirements level that can then be automatically built.

(2) **Generate tests from requirements that are meaningful to end-users.**
Suppose natural requirements from end-users are the only component dictating the software before it gets built. In that case, there is a broader scope for errors to be introduced into the resulting system. We need to ensure that end users gain confidence in the synthesised system by questioning whether it is working correctly. Testing systems in context is often the only way to ensure completeness of the requirements as even experienced software engineers would be unable to fully predict all downstream consequences of the deployed system. Within program synthesis, automated tests are generated to ensure that the software is working according to requirements. However, how do we make sure that these tests are meaningful to end users, such that they can trace errors back to the requirements that need to be amended?

(3) **Respond to dynamic requirements and environments.**
Applications are deployed once they satisfy end-user requirements. However, such requirements and application environments are dynamic. Changing requirements can cause misalignment between users' intents and the deployed applications [5]. Users need to trace the application performance back to requirements to understand areas of misalignment. Software development from requirements shortens development time therefore providing the opportunity for quicker iteration. If inconsistencies can be traced back to outdated or ambiguous requirements, this would obviate the need for manual fixes or patches and the software could be re-built correctly in real-time. Dynamic application environments can generate software failures and sub-optimal performance. Applications must use autonomous mechanisms to self-maintain and self-heal. These mechanisms must also enable transparent maintenance where end-users understand the failures and their causes but need not intervene.

In what follows, we outline a research agenda for requirements-driven EUSE and discuss associated risks and mitigation strategies.

## 2 RESEARCH AGENDA

### 2.1 Assist end users in expressing their requirements

Requirements elicitation has traditionally involved intermediaries between the user and the software. These might be people, such as UX researchers and software engineers, who translate the requirements of the user, or EUSE tools, which enable users to take actions in a particular programming language or environment. Both give rise to constraints around how and what the user can express.

In the new paradigm we are suggesting, designing whole apps from requirements allows the user to communicate directly with software, opening up the potential to realise the full nuance and complexity of the users' goals. Our vision exemplifies Sarkar's hypothesis that generative AI enables a "radical widening in scope and capability of end-user programming" [47].

Two challenges arise from this increased freedom to express requirements directly into software. The first is how to provide the required structure for users to succeed. Uncovering and defining requirements is hard to automate, as users often end up specifying requirements that are impractical, unnecessarily complex, or infeasible. The challenge is to guide users to express their requirements comprehensively and clearly enough to be built into software. The second challenge is helping the user to understand the possibilities and limitations of the software they can build. In other words, what areas of their workflow could they automate? Past work in EUSE by Ko et al. [28] called this a selection barrier: "finding what programming interfaces are available and which can be used to achieve a particular behaviour". Program synthesis alleviates the first part as it should match execution to requirements. Still, helping the user discover which behaviours are possible remains a challenge.

There are two strands of work in end-user software engineering that are directly relevant to the new paradigm we are proposing: programming by example and natural language programming.

Following years of research on programming by example [16], FlashFill, released in Excel 2013 [21], enabled widespread use. FlashFill automates string processing for end users by generalising from one or more examples. Building on this success, Generative AI shows tremendous capabilities to transform end user programming: a demo of Google's multimodal model, Gemini Pro [51], shows extraction of code to automate a web browsing task given just a screen recording of the user demonstrating the task.[6]

The second area of research is natural language adopted for end user programming. For instance, a concept spreadsheet, GridBook, demonstrated data analysis via natural language within the grid itself [50]. Following the success of GitHub Copilot since 2021 in empowering developers with code autocomplete, researchers adapted code-generating LLMs to generate spreadsheet calculations [33], with eventual impact in Excel Copilot. These developments show promise but their scope is limited to spreadsheet calculations. Requirements elicitation for whole app synthesis opens up new possibilities for what end users can build but also new challenges in an open environment beyond the spreadsheet. Alongside natural language, there is a great opportunity for end users to draw images or make short videos to express requirements.

---

[6]*Paige Bailey: Mind officially blown*, 22 February 2024.

Recent research explores the possibilities of translating natural language requirements into software applications across domains from creative endeavours to safety-critical contexts. For example, Vaithilingham et al. explore new capabilities that LLMs bring to design processes by developing an imagined scenario of an LLM-powered chat-based dialogue to elicit requirements for a video game [54]. Fakih et al. [18] built a "user-guided iterative pipeline" that integrates LLMs for programming in Industrial Control Systems. Their pipeline begins with an iterative loop between the user and the LLM to come up with a model that informs the subsequent code generation for Programmable Logic Controllers, which are used in industrial infrastructure applications. Wang et al. [55] design a process for translating high-level requirements specified in natural language into formal specifications for network configurations.

Facing the challenges we have introduced above, guiding the user to express their requirements in ways that can be built and understanding what it is possible to build in software, will involve addressing the following research questions. How do we build on research and techniques from HCI and Human Factors in this new paradigm? Can we leverage existing strategies for understanding user goals and preferences and redefine them in light of decreased constraints around time and resources when users themselves can be directly involved in creating software? LLMs might provide an opportunity to apply HCI methods to make requirements engineering more nuanced. For example, there is enormous potential to scale techniques that are currently time- and human-labour-intensive, such as ethnographic work and semi-structured interviews to develop an understanding of a user's context and needs.

Some examples of how requirements elicitation could be simplified using LLMs include: quicker iteration with users for earlier and more granular feedback on how their goals can be built; opportunity for post-deployment analysis and adjustments to the product; and using the utility of the Wizard of Oz technique [44] as inspiration to develop low-cost, quickly produced prototypes using LLMs to iterate on with users. How might these and other techniques be built into interfaces for requirements capture?

What do we miss when we lose the aspect of observing users executing their process? Observations are valuable input when generating requirements as sometimes they reveal tacit knowledge or subconscious priorities that a user would not consciously articulate. How can we build this into a process of direct interaction with software through requirements, for example leveraging programming by demonstration or multi-modal requirements elicitation?

How can we draw upon work from human factors to design workflows for producing software directly from requirements in ways that do not distract the user from their core goals or lead to loss of productivity in the ways explored by Simkute et al. [49]?

Moreover, what would a system look like from the perspectives of different domain-expert end-users? Could multiple versions of a system be built and then put together via iterative discussions between end-users about their requirements?

## 2.2 Generate tests from requirements that are meaningful to end-users

How can an end user tell whether their written requirements specify a whole app that meets their needs? After all, since the 1970s, critics of natural language specifications have pointed out the potential for ambiguity, even nonsense [17]. Even if there is no ambiguity, how can the end user tell whether the delivered app actually meets its requirements? LLMs are prone to hallucinate faulty code.

A professional software engineering team writes a suite of tests, such as input/output examples, to answer these questions. During requirements elicitation and code development, the team can:

(1) test its understanding of potentially ambiguous requirements, leading to progressive repair of the requirements, and
(2) test whether the code of the app meets the requirements, leading to progressive repair of the code.

Techniques for code generation of short function bodies from natural language descriptions already automate step (2). The HumanEval [14] and Mostly Basic Programming Problems (MBPP) [3] benchmarks both consist of a set of function specifications in natural language accompanied by a set of unit tests. Reflexion [48] achieved state of the art performance on HumanEval using a prompt chaining method that generates code, applies tests, reflects on error messages produced, and then repairs the errors discovered.

So, how can we help end users follow good practice and generate a comprehensive test suite to perform the steps above: (1) to test and repair requirements and (2) to test and repair code?

Helping end users write tests for their programs is a longstanding challenge in EUSE. End-user programming, such as formulas in spreadsheets, has been extremely error-prone. For example, Panko [42] found there is a high probability of error affecting the bottom lines of any substantial spreadsheet. Despite the prevalence of consequential errors, it has been hard to get end users to write tests. We face the characteristic challenge for EUSE that "the user probably has little expertise or even interest in software engineering" [7].

One hypothesis for future research on steps (1) and (2) is that if testing can be meaningfully aligned with a user's end goals, they might be more inclined to engage with it. Testing an app versus specifying its purpose in requirements involve different ways of thinking. Trying to both lay the groundwork for an app as well as question it within the same set of requirements would likely fail. Thus, we propose a separate requirements elicitation process for debugging where the end user can suggest a set of requirements for testing in the form of how their software should behave, what must happen and what must not happen, and then automatic unit tests are designed from this through program synthesis. In this way, testing might be more meaningful to users as they would be able to relate it to their goals and system understanding.

Future research can build on previous findings from testing and debugging tools for spreadsheets and other forms of end user computing. These tools make details visible [8], entice the user [46], utilise natural problem solving interactions like asking why questions [27], provide user interfaces to inspect formulas [20], and help users think about details in context [15] (perhaps via simulation or prototyping of a narrow slice of the requirements).

Ideally, we can auto-generate tests from the requirements using LLMs [19, 39]. For example, CodeT [12] improves code generation performance by generating tests automatically from natural language descriptions to effectively select appropriate code from amongst the many possibilities suggested by an LLM. More radically, recent work [36] using the verified programming language

Dafny [31] considers how to generate functions equipped with formal pre-conditions and post-conditions from natural language descriptions, as well as loop invariants to enable verification.

Still, when synthesised tests or specifications fail and cannot be automatically repaired, we need methods to communicate the situation to the end user, and seek their help. Information from the test failure needs to be intelligible to the user in terms of the high level requirements, as opposed to low level details of the code.

## 2.3 Respond to dynamic requirements and environments

The deployment and maintenance of applications are the next steps in the EUSE process. The deployment phase installs the application in the end user's environment. Maintenance tasks are responsible for the monitoring and optimal functioning of the application once deployed. These complex decision-making processes involve different variables at the requirements, software architecture, infrastructure, and environment levels. When considering EUSE, Lieberman advocates for automatic application deployment and maintenance [32]. Such automation is challenging because application requirements and environments are dynamic, demanding proactive adaptation to guarantee optimal functioning despite changing goals, variable data, unexpected failures, and security threats, among other variables that emerge from the real world [11]. Current efforts towards autonomous deployment rely on continuous delivery, DevOps, and MLOps tools [41]. These are successful solutions for automatically executing deployment pipelines. However, creating such pipelines is mostly a manual task in which experts define how and where to install each piece of software based on end user technical and budget constraints. Autonomous maintenance requires systems that self-adapt according to requirements and environmental changes. Current approaches in autonomous computing [43, 57] are often limited. They usually optimise simple requirements [9, 30] and fail to model complex phenomena [56].

Integrating capabilities from LLMs into the deployment and maintenance of synthesised applications has the potential to support the automatic creation of deployment plans and autonomous responses from applications to changing requirements, variable data, failures, and unexpected behaviour. The EUSE community must address challenging research questions before this potential is realised. These challenges emerge from the nature of LLMs, which are non-deterministic, prone to hallucinations, and operate as black boxes [19, 39]. Deployment plans must be consistent throughout the application life cycle, but LLMs can generate different plans for the same application because of their non-deterministic responses. The following questions arise: to what extent do non-deterministic responses impact the application deployment plan? How to validate deployment plans before execution? How to handle stochasticity when executing deployment plans? And how to integrate deployment plans generated by LLMs with current tools (e.g., continuous delivery, DevOps, MLOps, etc.)? Prompt engineering or Retrieval-Augmented Generation [13] are promising research directions [10] as these use external data to make LLMs' outputs more consistent.

LLM hallucinations can cause application misbehaviour by injecting errors into software components, deployment plans, and maintenance decisions. There is already progress on automatically resolving a benchmark set of GitHub issues [23], including maintenance tasks. The holistic deployment and maintenance of applications still have the following open questions: to what extent can we use LLMs to build applications with critical requirements (e.g. in the healthcare domain)? How to handle hallucinations to generate software components and deployment and maintenance plans? How to identify faulty or outdated components generated by LLMs? How to identify causal relationships between requirements and components to enable preventive or corrective maintenance? How to use these causal relationships to support alignment analysis between users' requirements and applications? A promising research direction is the work on data-oriented architectures (DOAs) [11], which can enable causal analysis on systems data [40].

Deployment and maintenance processes must be transparent, automated, and interpretable to end users. LLMs offer new interfaces between applications and end users. But questions regarding explainability remain open: how to explain deployment and maintenance decisions made by LLMs to end users? Can the interaction between end-users, LLMs and other entities in the synthesis process (e.g. external data sources) improve such explainability? As discussed in section 2.1, techniques from HCI could support the development of system monitoring tools along these lines.

## 3 RISKS AND MITIGATIONS

Research on requirements-driven EUSE seeks AI-based algorithms to convert requirements into whole apps. In some cases, these algorithms could generate apps that are biased against certain demographic groups. *Algorithmic audit* is a "method of repeatedly querying an algorithm and observing its output to analyse the algorithm's opaque inner workings and possible external impact" [35]. Suppliers of these algorithms should assist in external algorithmic audits, and help automate the administration of these.

Responsible AI research has led to large companies issuing guidelines and checklists [2] while researchers examine their applicability in practice [45]. We will need guidelines for responsible requirements-based EUSE. The uncertain nature of Generative AI can inject threats into synthesised applications with critical requirements. Minimising these threats is key towards the realisation of our research vision. Based in part on hazard analysis of Codex [25], Khlaff [24] proposes the Operational Design Domain (ODD) model, originally developed to categorise Automated Driving Systems [52], to define concrete operational envelopes for AI-based systems. We should define a hierarchy of ODDs for requirements-driven EUSE to help audit and contain the operational risks.

## 4 CONCLUSIONS

Breakthroughs in LLMs have enabled a paradigm shift in the control users can have over the software development life cycle. Our vision of requirements-driven end-user software engineering is one in which end users can conceptualise software, test and deploy it entirely from requirements. We have looked at the limits to how far requirements can go in dictating system behaviour including non-functional requirements which need to be automated but still communicated clearly to the user. We propose an EUSE research agenda with three key areas of focus to make this vision a reality.